\font\bbf       =cmbx12
\overfullrule 0pt
\vsize=9truein
\hsize=6.5truein

\def \A {{\hbar \omega^3 \over 2\pi^2 c^3}}

\def \Is {$I_*$}
\def \Isp {$I'_*$}
\def \It	{$I_{\tau}$}
\def \Itp	{$I'_{\tau}$}

\def \Itz	{$I_{\tau=0}$}
\def \Isl {$I_{*,L}$}
\def \Itl	{$I_{\tau,L}$}

\def \w {\omega}

\def \ew {\eta(\omega)}

\def \Fszp	{{\bf f}_*^{zp}}

\def \t {\tau}

\def \ptg {$(c^2/g, 0, 0)$}
\bigskip
\centerline{\bbf Gravity and the Quantum Vacuum Inertia Hypothesis}
\centerline{\bbf I. Formalized Groundwork for Extension to Gravity}
\bigskip
\centerline{Alfonso Rueda}
\centerline{\it Department of Electrical Engineering \& Department of Physics, ECS
Building}
\centerline{\it California State University, 1250 Bellflower Blvd.,
Long Beach, CA 90840}
\centerline{arueda@csulb.edu}
\bigskip
\centerline{Bernard Haisch and Roh Tung}
\centerline{\it California Institute for Physics \& Astrophysics}
\centerline{\it 366 Cambridge Ave., Palo Alto, CA 94306}
\centerline{haisch@calphysics.org, tung@calphysics.org}
\bigskip
\centerline{\bf Abstract}
\bigskip
It has been shown [1,2] that the
electromagnetic quantum vacuum makes a contribution to the inertial mass, $m_i$, in
the sense that at least part of the inertial force of opposition to acceleration, or
inertia reaction force, springs from the electromagnetic quantum vacuum (see also
[3] for an earlier attempt).  Specifically, in the previously cited work, the
properties of the electromagnetic quantum vacuum as experienced in a Rindler constant
acceleration frame were investigated, and the existence of an energy-momentum flux
was discovered which, for convenience, we call the Rindler flux (RF).  The RF, and its
relative, Unruh-Davies radiation, both stem from event-horizon effects in
accelerating reference frames. The force of radiation pressure produced by the RF
proves to be proportional to the acceleration of the reference frame, which leads to
the hypothesis that at least part of the inertia of an object should be due to the
individual and collective interaction of its quarks and electrons with the RF. We call
this the {\it quantum vacuum inertia hypothesis}. We demonstrate that this approach to
inertia is consistent with general relativity (GR) and that it answers a fundamental
question left open within GR, viz. is there a physical mechanism that generates the 
reaction force known as weight when a specific non-geodesic motion is imposed on an
object? Or put another way, while geometrodynamics dictates the spacetime metric and
thus specifies geodesics, is there an identifiable mechanism for enforcing the motion
of freely-falling bodies along geodesic trajectories? The quantum vacuum inertia
hypothesis provides such a mechanism, since by assuming the Einstein principle of
local Lorentz-invariance (LLI), we can immediately show that the same RF arises due
to curved spacetime geometry as for acceleration in flat spacetime. Thus the
previously derived expression for the inertial mass contribution from the
electromagnetic quantum vacuum field is exactly equal to the corresponding
contribution to the gravitational mass,
$m_g$. Therefore, within the electromagnetic quantum vacuum viewpoint
proposed in [1,2], the Newtonian weak equivalence principle,
$m_i=m_g$, ensues in a straightforward manner. In the weak field limit it can then
also be shown, by means of a simple argument from potential theory, that because of
geometrical reasons the Newtonian gravitational force law must exactly follow. This
elementary analysis however does not pin down the exact form of the gravitational
theory that is required but only that it should be a theory of the metric type, i.e.,
a theory like Einstein's GR that can be interpreted as curvature of spacetime.  While
the present analysis shows that our previous quantum vacuum inertial mass analysis is
consistent with GR, the extension of these two analyses to components of
the quantum vacuum other than the electromagnetic component, i.e. the strong and weak
vacua, remains to be done.
\bigskip\noindent
{1. INTRODUCTION}
\bigskip
Using the semiclassical representation of the electromagnetic quantum vacuum embodied
in Stochastic Electrodynamics (SED), it has been shown that a contribution to the
inertial mass, $m_i$, of an object must result from the interactions of the quantum
vacuum with the electromagnetically-interacting particles (quarks and electrons)
comprising that object [1,2].
Specifically, the
properties of the electromagnetic quantum vacuum as measured in a Rindler constant
acceleration frame were investigated, and the existence of an energy-momentum flux
was discovered which, for convenience, we now call the Rindler flux (RF).  The RF, and
its relative, Unruh-Davies radiation, both stem from event-horizon effects in
accelerating reference frames.  Event horizons create an asymmetry in the quantum
vacuum radiation pattern.

The force of radiation pressure on a massive object produced by the RF proves to be
proportional to the acceleration of the reference frame attached to the object.
This leads naturally to the hypothesis that at least part of the inertia of an object
is due to the acceleration-dependent drag force that results from
individual and collective interactions of the quarks and electrons in the object with
the RF. For simplicity of reference, we refer to this concept, and an earlier
derivation of a similar result using a completely different approach (involving a
perturbation technique due to Einstein and Hopf on an accelerating Planck oscillator
[3]), as the {\it quantum vacuum inertia hypothesis}.

SED is a theory that includes the effects of the electromagnetic
quantum vacuum in physics by adding to ordinary Lorentzian classical electrodynamics a
random fluctuating electromagnetic background constrained to be homogeneous and
isotropic and to look exactly the same in every Lorentz inertial frame of
reference [4,5]. This replaces the zero homogeneous background
of ordinary classical electrodynamics. It is essential that this background not
change the laws of physics when exchanging one inertial reference system for another.
This translates into the requirement that this random electromagnetic background
must have a Lorentz invariant energy density spectrum. The only random electromagnetic
background with this property is one whose spectral energy density,
$\rho(\w)$, is proportional to the cube of the frequency, $\rho(\w) d\w
\sim \w^3 d\w$. This is the case if the energy per mode is $\hbar
\w/2$ where
$\w$ is the angular frequency. (The $\hbar \w/2$ energy per mode is of course
also the minimum energy of the analog of an
electromagnetic field mode: a harmonic oscillator.) The spectral energy
density required for Lorentz invariance is thus identical to the spectral energy
density of the zero-point field of ordinary quantum theory. For most purposes,
including the present one, the zero-point field of SED may be identified with the
electromagnetic quantum vacuum. However SED is essentially a classical theory since
it presupposes only ordinary classical electrodynamics and hence SED also presupposes
special relativity (SR).

According to the weak equivalence principle (WEP) of Newton and Galileo, inertial
mass is equal to gravitational mass, $m_i=m_g$. If the quantum vacuum inertia
hypothesis is correct, a very similar mechanism involving the quantum vacuum should
also account for gravitational mass. This novel result, restricted for the time being
to the electromagnetic vacuum component, is precisely what we show in
$\S 2$ by means of formal but simple and straightforward arguments requiring physical
assumptions that are uncontroversial and widely accepted in theoretical physics.
In $\S 3$ the consistency of this argument with so-called metric theories of gravity
(i.e. those theories characterized by spacetime curvature) is exhibited. In addition
to the metric theory, {\it par excellance}, Einstein's GR, there is the Brans-Dicke
theory and other less well known ones, briefly discussed by Will [6].
$\S4$ briefly discusses a non-metric theory.

Nothing in our approach points to any new discriminants among the various metric
theories. Nevertheless, our quantum vacuum approach to gravitational mass will be
shown to be entirely consistent with the standard version of GR. Next, in $\S 5$, we
take advantage of geometrical symmetries and present a short argument from standard
potential theory to show that in the weak field limit a Newtonian inverse square
force must result from our approach.

A new perspective on the origin of weight is presented in $\S 6$.
In $\S 7$ we discuss an energetics aspect, related to the derivation presented
herein, and resolve an apparent paradox. 
A brief discussion on the nature of the
gravitational field follows in
$\S 8$, but we infer that the present development of our approach does not provide any
deeper or more fundamental insight than GR itself into the ability of matter to bend
spacetime. We present conclusions in $\S 9$. A full development within GR is left for
the accompanying article [7].

\bigskip\noindent 2. 	ON WHY THE ELECTROMAGNETIC VACUUM CONTRIBUTION TO
GRAVITATIONAL MASS IS EXACTLY THE SAME AS FOR INERTIAL MASS
\bigskip
Table 1 compares the quantum vacuum inertia hypothesis with the standard view
on mass. We intend to show -- in this and a companion paper [7] -- not only that the
quantum vacuum inertia hypothesis is consistent with GR, but that it answers an
outstanding question regarding a possible physical origin of the force manifesting as
weight. We also intend to show that just as it becomes possible to identify a
physical process underlying the {\bf f}=m{\bf a} postulate of Newtonian
mechanics (as well as its extension to SR [1,2]), it is possible to identify a
parallel physical process underlying the weak equivalence principle, $m_i=m_g$, viz.
interaction of matter with the RF.

Within the standard theoretical framework of GR and related theories, the equality (or
proportionality) of inertial mass to gravitational mass has to be assumed. It remains
unexplained. As correctly stated by Rindler [8], ``the proportionality of inertial
and gravitational mass for different materials is really a very mysterious fact.''
However here we show that -- at least within the present restriction to
electromagnetism -- the quantum vacuum inertia hypothesis leads naturally and
inevitably to this equality. The interaction between the electromagnetic quantum
vacuum and the electromagnetically-interacting particles constituting any physical
object (quarks and electrons) is identical for the two situations of acceleration
with respect to constant velocity inertial frames or remaining fixed above some
gravitating body with respect to freely-falling local inertial frames.

A related theoretical lucuna involves the origin of the force which manifests itself
as weight. Within GR theory one can only state that deviation from geodesic motion
results in a force which must be an inertia reaction force. We propose that it is
possible in principle to identify a mechanism which
generates such an inertia reaction force, and that in curved spacetime it
acts in the same way as acceleration does in flat spacetime.

Begin by considering a macroscopic, massive, gravitating object, $W$, which is fixed
in space and for simplicity we assume to be solid, of constant density, and spherical
with a radius
$R$, e.g. a planet-like object. At a distance $r>>R$ from the center of $W$ there is a
small object,
$w$, that for our purposes we may regard as a point-like test particle. A constant
force
$\bf f$ is exerted by an external agent that prevents the small body $w$ from falling
into the gravitational potential of $W$ and thereby maintains $w$ at a fixed point in
space above the surface of $W$. Experience tells us that when the force {\bf f} is
removed, $w$ will instantaneously start to move toward $W$ with an acceleration
{\bf g} and then continue freely falling toward $W$.

Next we consider a freely falling {\it local} inertial frame \Is \ (in the
customary sense given to such a local frame [6]) that is instantaneously at
rest with respect to $w$. 
At $w$ proper time $\t$, that we select to be
$\t=0$, object $w$ is instantaneously at rest at the point \ptg \ of the \Is \
frame. The $x$-axis of that frame goes in the direction from $W$ to $w$ and, since
the frame is freely falling toward $W$, at $\t = 0$ object $w$ appears accelerated
in \Is \ in the $x$-direction and with an acceleration ${\bf g}_w=\hat {x}g$. As
argued below,
$w$ is performing a uniformly-accelerated motion, i.e. a motion with a constant proper
acceleration ${\bf g}_w$ as observed from any neighboring instantaneously comoving
(local) inertial frame. In this respect we introduce an infinite collection of local
inertial frames \It, with axes parallel to those of \Is \ and with a common $x$-axis
which is that of \Is . Let $w$ be instantaneously at rest and co-moving
with the frame \It \ at $w$ proper time $\t$.
So the $\t$ parameter representing the $w$ proper time also serves to
parametrize this infinite collection of (local) inertial frames.
Clearly then, \Is \ is the member of the collection with $\t =0$, so that
$I_*=I_{\t=0}$.
At the point in time of coincidence with a given \It,
$w$ is found momentarly at rest at the \ptg \ point of the \It \ frame. We
select also the times in the (local) inertial frames to be $t_{\t}$ and such that
$t_{\t}=0$ at the moment of coincidence when $w$ is instantaneously at rest in \It \
and at the aforementioned \ptg \ point of \It.
Clearly as $I_{\t=0}=I_*$ then $t_*=0$ when $\t=0$. All the frames in the collection
are freely falling toward $W$ and when any one of them is instantaneously at rest
with $w$ it is instantaneously falling with acceleration
${\bf g}=-{\bf g}_w=-\hat{x}g$ with respect to $w$ in the direction of $W$.
It is not difficult to realize that $w$ appears in those frames as uniformly
accelerated and hence performing a hyperbolic motion with constant proper
acceleration
${\bf g}_w$.

This situation is equivalent to that of an object $w$ accelerating with respect to
an ensemble of \It \ reference frames in the absence of gravity. In that
situation, the concept of the ensemble of inertial frames,
\It, each with an infinitesimally greater velocity (for the case of positive
acceleration) than the last, and each coinciding instantaneously with an accelerating
$w$ is not difficult to picture. But how does one picture the analogous ensemble for
$w$ held stationary with respect to a gravitating body?

We are free to bring reference frames into existence at will. Imagine bringing a
reference frame into existence at time $\tau=0$ directly adjacent to $w$, but whereas
$w$ is fixed at a specific point above $W$, we let the newly created reference frame
immediately begin free-falling toward $W$. We immediately create a replacement
reference frame directly adjacent to $w$ and let it drop, and so on. The
ensemble of freely-falling local inertial frames bear the same relation to $w$ and to
each other as do the extended \It \ inertial frames used in the case of true
acceleration of
$w$.

For convenience we introduce a special frame of reference, $S$, whose $x$-axis
coincides with those of the \It \ frames, including of course \Is, and whose $y$-axis
and
$z$-axis are parallel to those of \It \ and \Is. This frame $S$ stays collocated with 
$w$ which is positioned at the \ptg \ point of the $S$ frame. For \Is \ (and for the
\It \ frames) the frame $S$ appears as accelerated with the uniform acceleration ${\bf
g}_w$ of its point \ptg . We will assume that the frame $S$ is rigid. If so, the
accelerations of points of $S$ sufficiently separated from the $w$ point \ptg \ are
not going to be the same as that of \ptg. This is not a concern however since we will
only need in all frames (\It, \Is, \ and $S$) to consider points in a sufficiently
small neighborhood of the \ptg \ point of each frame

The collection of frames, \It, as well as \Is \ and  $S$, correspond exactly to
the set of frames introduced in [1]. The only differences are, first, that now
they are all local, in the sense that they are only well defined for regions in the
neighborhood of their respective \ptg \ space points; and second, that now \Is \ and
the \It \ frames are all considered to be freely falling toward $W$ and the $S$ frame
is fixed with respect to $W$. Similarly to [1], the $S$ frame may again be
considered to be, relative to the viewpoint of \Is, a {\it Rindler noninertial
frame}. The laboratory frame \Is \ we now call the {\it Einstein laboratory frame},
since now the ``laboratory'' is local and freely falling. We call the collection of
inertial frames \It \ the Boyer family of frames as he was the first to introduce
them in SED [9].

The relativity principle as formulated by Einstein when proposing SR states that ``all
inertial frames are totally equivalent for the performance of all physical
experiments.''[6] Before applying this principle to the freely-falling frames \Is
\ and \It \ that we have defined above it is necessary to draw a distinction between
these frames and inertial frames that are far away from any gravitating body, such as
$W$. The free-fall trajectories, i.e. the geodesics, in the vicinity of any
gravitating body, $W$, cannot be parallel over any arbitrary distance owing to the
fact that $W$ must be of finite size. 
This means that the principle of relativity can only be applied locally. This was
precisely the limitation that Einstein had to put on his infinitely-extended Lorentz
inertial frames of SR when starting to construct GR [6,8,10].

We adopt the principle of local Lorentz invariance, (LLI) which can be stated,
following Will [6], as ``the outcome of a local nongravitational test experiment is
independent of the velocity of the freely falling apparatus.'' A non-gravitational
test experiment is one for which self-gravitating effects can be neglected. We also
adopt the assumption of space and time uniformity, which we call the uniformity
assumption (UA) and which states that the laws of physics are the same at any time or
place within the Universe. Again, following Will [6] this can be stated as ``the
outcome of any local nongravitational test experiment is independent of where and
when in the universe it is performed.'' We do not concern ourselves with physical
cosmological theories that in one way or another violate UA, e.g. because they involve
spatial or temporal changes in fundamental constants [11].

Locally, the freely falling local Lorentz frames which we now designate with a
subscript
$L$ --- \Isl \ and \Itl \ ---
are entirely equivalent to the \Is \ and \It \ extended frames of [1]. The
free-falling Lorentz frame \Itl \ locally is exactly the same as the extended \It.
Invoking the LLI principle we can then immediately conclude that the
electromagnetic zero-point field, or electromagnetic quantum vacuum, that can be
associated with
\Itl \ must be the same as that associated with  \It. From the viewpoint of the
local Lorentz frames \Isl \ and \Itl \ the body $w$ is undergoing uniform
acceleration and therefore for the same reasons as presented in [1] an
acceleration-dependent drag force arises.

{\it These formal arguments demonstrate that the analyses of of [1,2] which found
the existence of a RF in an accelerating reference frame translate and
correspond exactly to a reference frame fixed above a gravitating body.}
In the same manner that light rays are deviated from straight-line propagation by a
massive gravitating body $W$, the other forms of electromagnetic radiation, including
the electromagnetic zero-point field rays (in the SED approximation) are also
deviated from straight-line propagation. Not surprisingly this creates an anisotropy
in the otherwise isotropic electromagnetic quantum vacuum. This is the origin of the
RF in the gravitational case.

In [1] we interpreted the drag force exerted by the RF as the inertia
reaction force of an object that is being forced to accelerate through the
electromagnetic zero-point field. Accordingly, in the present situation, the
associated nonrelativistic form of the inertia reaction force should be

$$\Fszp=-m_i {\bf g}_w \eqno(1)
$$
where ${\bf g}_w$ is the acceleration with which $w$ appears in the local inertial
frame \Is. As shown in [1] the coefficient $m_i$ is

$$m_i= \left[ {V_0 \over c^2} \int \ew \A d\w \right] 
\eqno(2)$$
where $V_0$ is the proper volume of the object, $c$ is the speed of light, $\hbar$ is
Planck's constant divided by $2 \pi$ and $\ew$, where $0 \le \ew \le 1$, is a
function that spectralwise represents the relative strength of the interaction between
the zero-point field and the massive object which acts
to oppose the acceleration. If the object is just a single particle, the spectral
profile of
$\ew$ will characterize the electromagnetically-interacting particle. It can also
characterize a much more extended object, i.e. a macroscopic object, but then the
$\ew$ will have much more structure (in frequency). We should expect different shapes
for the electron, a given quark, a composite particle like the proton, a molecule, a
homogeneous dust grain or a homogeneous macroscopic body. In the last case the $\ew$
becomes a complicated spectral opacity function that must extend to extremely high
frequencies such as those characterizing the Compton frequency of the electron and
even beyond.

Now, however, what appears as inertial
mass,
$m_i$, to the observer in the local \Isl \ frame is of course what corresponds to
gravitational mass, $m_g$, and it must therefore be the case that

$$m_g= \left[ {V_0 \over c^2} \int \ew \A d\w \right] .
\eqno(3)$$
As done in [1], Appendix B, it can be shown that the right hand side indeed
represents the energy of the electromagnetic quantum vacuum enclosed within the
object's volume and able to interact with the object as manifested by the $\ew$
coupling function. A more thorough, fully covariant development can also be
implemented to show that the force expression of eqn. (1) can be extended to the
relativistic form of the inertia reaction force as in [1] , Appendix D. (This
development also served to obtain the final form of $m_i$ given above in eqn. (3)
eliminating a spurious 4/3 factor.)
\footnote{*}{We use this opportunity to correct a minor transcription error that
appeared in the printed version of Ref. [1]. In Appendix D, p. 1100, the minus sign in
eqn. (D8) is wrong. It should read
$$\eta^{\nu} \eta_{\nu}=1 \eqno({\rm D}8)
$$
and the corresponding signature signs in the line just above eqn. (D8) are the
opposite of what was written and should instead read $(+ - - \ - )$.
}
Summarizing what we have shown in this section is that if a force {\bf f} is applied
to the $w$ body just large enough to prevent it from falling toward the body $W$,
then in the non-relativistic case that force is given by

$${\bf f}=-m {\bf g} \eqno(4)
$$
where we have dropped the nonessential subscripts $i$ and $g$ and superscripts,
because it is now clear that $m_i=m_g=m$ follows from the quantum vacuum inertia
hypothesis.

\bigskip\noindent
3. CONSISTENCY WITH EINSTEIN'S GENERAL RELATIVITY
\bigskip
The statement that $m_i=m_g$ constitutes the weak equivalence principle (WEP). Its
origin goes back to Galileo and Newton, but it now appears, as shown in the previous
section, that this principle is a natural consequence of the quantum vacuum inertia
hypothesis. The strong equivalence principle (SEP) of Einstein consists of the WEP
together with LLI and the UA. Since the quantum vacuum inertia hypothesis and its
extension to gravity allow us to obtain the WEP assuming LLI and the UA, this
approach is consistent with all theories that are derived from the SEP. In addition
to GR, the Brans-Dicke theory is derived from SEP as are other
other lesser known theories [6], all distinguished from each other by
various particular assumptions.

All theories that assume the SEP are called {\it metric theories}. They are
characterized by the fact that they contemplate a bending of spacetime associated
with the presence of matter. Two important consequences of the LLI-WEP-UA combination
are that light bends in the presence of matter and that there is a gravitational
Doppler shift. Since the quantum vacuum inertia hypothesis is consistent with this
same combination, it would also require that light bends in the presence of
gravitational fields. This can, of course, be interpreted as a change in spacetime
geometry, the standard interpretation of GR.

\bigskip\noindent
4. A RECENT ALTERNATIVE VACUUM APPROACH TO GRAVITY
\bigskip\noindent

The idea that the vacuum is ultimately responsible for gravitation is not new. It
goes back to a proposal of Sakharov [12] based on the work of Zeldovich [13] in
which a connection is drawn between Hilbert-Einstein action and the quantum vacuum.
This leads to a view of gravity as ``a metric elasticity of space'' (see Misner,
Thorne and Wheeler [14] for a succinct review of this concept). Following Sakharov's
idea and using the techniques of SED, Puthoff proposed that gravity could be
construed as a form of van der Waals force [15]. Although interesting and
stimulating in some respects, Puthoff's attempt to derive a Newtonian inverse square
force of gravity proves to be unsuccessful [16,17,18,19].

An alternative approach has recently been developed by Puthoff [20] that is based on
earlier work of Dicke [21] and of Wilson [22]: a polarizable vacuum model of
gravitation. In this representation, gravitation comes from an effect by massive
bodies on both the permittivity, $\epsilon_0$, and the permeability, $\mu_0$, of the
vacuum and thus on the velocity of light in the presence of matter.

This is clearly an alternative theory to GR since it does not involve
curvature of spacetime. So it is not a metric theory of gravity. It can be seen that
it does not entail the SEP of Einstein. On the other hand, since spacetime curvature
is by definition inferred from light propagation in relativity theory, the polarizable
vacuum gravitation model may be labelled a pseudo-metric theory of gravitation since
the effect of variation in the dielectric properties of the vacuum by massive objects
on light propagation are approximately equivalent to GR spacetime curvature as long
as the fields are sufficiently weak.
 
In the weak-field limit, the polarizable vacuum
model of gravitation duplicates the results of GR, including the classic tests
(gravitational redshift, bending of light near the Sun, advance of the perihelion of
Mercury). Differences appear in the strong-field regime, which should lead to
interesting tests.

\bigskip\noindent
5. DERIVATION OF NEWTON'S LAW OF GRAVITATION
\bigskip\noindent

Our approach allows us to derive a Newtonian form of gravitation in the weak-field
limit based on the quantum vacuum inertia hypothesis, local Lorentz invariance and
geometrical considerations. In [1] we showed how an asymmetry that appears in
the radiation pattern of the electromagnetic quantum vacuum, when viewed from an
accelerating reference frame, leads to the appearance of a non-zero
momentum flux, the Rindler Flux (RF), opposing any accelerating
object. Individual and collective interaction between the
electromagnetically-interacting particles (quarks and electrons) comprising a
material object and the RF generates a reaction force that may be identified with the
inertia reaction force as it has the right form for all velocity regimes. In
particular in the low velocity limit it is exactly proportional to the acceleration
of the object.

In $\S 2$ we showed from formal arguments based on the principle of local Lorentz
invariance (LLI) that an exactly equivalent force must originate when an object is
effectively accelerated with respect to free-falling Lorentz frames by virtue of
being held fixed above a gravitating body, $W$. We infer that the presence of a
gravitating body must distort the electromagnetic quantum vacuum in exactly the same
way at any given point as would the process of acceleration such that ${\bf a}=-{\bf
g}$. Simple geometrical arguments [23] now suffice to show that the gravitational
force can only be the Newton inverse-square law with distance (in the weak field
limit).

It has been shown ($\S2$) that (outside $W$) the {\bf g} field of eqn. (4) generated
by
$W$ has to be central, i.e., centrally distributed with spherical symmetry around
$W$. It has to be radial, with its vectorial direction parallel to the corresponding
radius vector {\bf r} originating at the center of mass of $W$ where we locate the
origin of coordinates. The spherical symmetry implies that {\bf g} is radial in the
direction
$-\hat{\bf r}$, and depends only on the $r$-coordinate.

The field is clearly generated by mass. A simple symmetry argument, here omitted for
brevity [24], shows that indeed if we scale the mass
$M$  of
$W$ by a factor $\alpha$ then the resulting {\bf g} must be of the
form $\alpha {\bf g}$. If $M$ goes to zero, {\bf g} disappears.
The mass $M$ must be the source of field
lines of {\bf g}, and these field lines can be discontinuous only where mass is
present. The field lines can be neither generated nor destroyed in free space.  Since
{\bf g} is the force on a unit mass, we must expect that {\bf g} behaves as a vector,
and specifically that {\bf g} follows the laws of vector addition. Namely, if two
masses
$M_1$ and $M_2$ in the vicinity of each other generate fields ${\bf g}_1$ and ${\bf
g}_2$ respectively, the resulting {\bf g} at any given point in space should be the
linear superposition

$${\bf g}={\bf g}_1+{\bf g}_2 .
 \eqno(5)
$$
Finally, from the argument that leads to eqn. (4) we can see
that the field {\bf g} must be unbounded, extending essentially to infinity.

With all of these considerations, clearly the lines of {\bf g} must obey the
continuity property outside $W$. If there is no mass present inside a volume, $V$,
enclosed by a surface $S$, we expect that

$$0=\oint_{S(V)} {\bf g} \cdot {\bf n} \ dS = \int_V \nabla \cdot {\bf g} \ dV ,
\eqno(6)
$$
but since $V$ is arbitrary this tells us that outside the massive body $W$

$$\nabla \cdot {\bf g}=0 , \ \ \ \ r > R \ . \eqno(7)
$$
In the presence of our single massive body, $W$, but outside that body,
{\bf g} is radial from the center of mass, and therefore from eqn. (7) we conclude
that

$${\bf g} \sim -\hat{\bf r} {1 \over r^2}  \eqno(8)
$$
where the restriction $r>R$ is hereafter understood.
Since the field {\bf g} is also proportional to the mass $M$ which is its origin, we
conclude that {\bf g} must be of the form

$${\bf g} = -\hat{\bf r} {GM \over r^2} . \eqno(9)
$$
where $G$ is a proportionality constant and from eqn. (4) we have that
$${\bf f} = -\hat{\bf r} {GMm \over r^2} . \eqno(10)
$$
which is Newton's law of gravitation. It is remarkable that after finding
the central and radial charater of {\bf g} by means of the vacuum approach of
our quantum vacuum inertia hypothesis, one can immediately obtain Newtonian
gravitation, an endeavour keenly but unsuccessfully pursued for quite some time from
the viewpoint of the vacuum fields [12,13,14] and in particular of SED
[15,16,17,18,19]. In this last case (SED), it was proposed that gravity was a force of
the van der Waals form, a view which has been shown to be unsuccessful [19].

\bigskip\noindent
6. ON THE ORIGIN OF WEIGHT
\bigskip

We have established that the
quantum vacuum inertia hypothesis leads to a force in eqn. (10) which adequately
explains the origin of weight in a Newtonian view of gravitation. How is this
consistent with the geometrodynamic view of GR? Geometrodynamics specifies the
effect of matter and energy on an assumed pliable spacetime metric. That defines the
geodesics which light rays and freely-falling objects will follow. However there is
nothing in geometrodynamics that points to the origin of the inertia reaction
force when geodesic motion is prevented, which manifests in special circumstances as
weight.

Geometrodynamics merely assumes that deviation from geodesic motion results in
inertial forces. That is, in fact, true, but as stated is devoid of any
physical insight as discussed in detail in [25]. What we have shown above is that an
identical asymmetry in the quantum vacuum radiation pattern will arise due to either
true acceleration or to effects on light propagation by the presence of gravitating
matter. That identical asymmetry leads to identical non-zero RFs in the
electromagnetic zero-point field, which generate identical forces. In the case of
true acceleration, the resulting force is the inertia reaction force. In the case of
being held stationary in a non-Minkowski metric, the resulting force is the weight,
which is also the enforcer of geodesic motion for freely-falling objects.

\bigskip\noindent
7. A PHANTOM ANOMALY
\bigskip

Following the reasoning of $\S 6$ one would conclude that at every point of fixed $r$
above $W$ there is an inflowing RF. This would seem to imply a continuous energy flux
toward and into $W$, an apparently paradoxical situation consisting of quantum
vacuum-originating energy streaming toward $W$ from all directions in space.

The resolution to this apparent paradox comes from the realization that we are
naively summing energy flows from incompatible reference frames. Two observers at the
same
$r$ but 180 degrees apart will report energy flows in opposite directions. These
energy flows are locally true, but they come from oppositely-directed reference
frames. Only in a freely-falling frame in which all of the energy fluxes could
simultaneously be properly defined could we legitimately simply sum them up. However
it is mathematically improper and non-physical to add up over vectors defined in
different reference frames, which moreover in the present case can be said to even be
incompatible.

For the sake of definiteness, let us locate
ourselves and $w$ not very far away from the surface of $W$, and for simplicity let
$W$ be perfectly spherical with radius $R$ and homogeneous in density. Let the test
body $w$ be, as before, much smaller than $W$ and assume that it is prevented from
falling toward $W$ by a supporting force {\bf f} as in eqn. (4).

We need only realize two facts. The analysis leading to the RF that
appears for $w$ is only consistent for a given freely-falling LLI frame or at most
for the Boyer family of frames \It, where \Itz=\Is, along the particular straight
radius vector going from $W$ to $w$. (For simplicity of notation we drop the $L$-index
subscript that indicates the locality of the frames as there is no source of confusion
at this point.)

For example, consider a small body $w'$ (resting at the \ptg \ point of $S'$)
and associated set of frames \Itp \ and \Isp \ falling freely toward
$W$ and defined exactly the same as \It \ and \Is \ were defined with respect to $w$
but whose radius vector of free fall toward $W$ is along the widely different
direction that goes from $W$ to $w'$. As occurs with the observers of unprimed frames
\It \ and \Is, observers in the primed frames \Itp \ and \Isp \ also claim there
appears a RF in $S'$ along the radial direction or $x'$-axis and
hitting the body $w'$ in the direction pointing toward $W$. But for the primed frame,
the unprimed frame's direction of fall toward $W$ is not a direction of any net
RF. The reverse conclusion also holds. For the unprimed frame there is
no net RF along the radial direction ($x'$-axis) of the primed frame.

Therefore all this shows that those RFs only can be inferred as such
from a selectively restricted class of LLI frames. The only way we could produce a
consistent conclusion about this odd thermodynamics situation is if we could find a
single freely-falling LLI frame from which we could simultaneously make a consistent
conclusion about the nature of all such RFs, i.e. a frame from which we can define
simultaneously and hence be able to add up over all those different RFs. Fortunately
there exists such a frame.

Consider $W$ as the sphere of the Earth. As one goes more and more deeply inside, the
strength of {\bf g} decreases. So freely-falling frames ever deeper within the Earth
fall with ever smaller accelerations, and at the exact center of mass of the Earth
{\bf g} is exactly zero. We can actually quantify this in full precision since we
have already discovered that the gravitational force must go as $r^{-2}$.

There is one single freely-falling LLI frame that is a member of all radial families
of freely-falling frames and that is the LLI frame exactly at the center of the Earth
that we shall call $C$. From that frame $C$ we can observe and draw consistent
conclusions about the nature of all the RFs along all possible radial
directions toward the center of $W$. But $C$ is not only freely falling
with exactly zero acceleration, but also has always been at rest at the center of the
Earth. Furthermore conclusions made from this freely-falling frame can be universal
since it has zero acceleration and, at least from this limited perspective, can be
extended indefinitely, i.e. need not be strictly local.

Then for that frame, $C$, we can consistently look in all possible radial directions.
Consider the direction from the center of $W$ to $w$. Since $w$ is at a fixed
distance from the center of $W$ and this is fixed and not moving with respect to an
observer in $C$, there cannot be any RF due to quantum vacuum radiation of
$C$ impinging on $w$, because $w$ does not appear accelerated as viewed from $C$. So
the RF disappears along the particular radial direction. But for that
matter, it disappears along all radial directions from the center of $W$ and the
paradox is resolved. The only single frame from which a consistent conclusion
referring to all RFs along all possible radial directions from the center
of $W$ can be drawn yields that such RFs exactly vanish. The observer in
$C$ does not infer any net radiation influx toward $W$ from any direction in space.

We also note that any observer in orbit around $W$ will detect no RF. Such
an observer will detect a RF should he attempt to change orbits, i.e.
deviate from geodesic motion. Indeed, he must apply a force to overcome the effect of
the RF when he attempts to change orbits.

\bigskip\noindent
8. DISCUSSION
\bigskip\noindent

In light of what has been proposed herein, what can we
elucidate about the nature of the gravitational field? In the low fields
and low velocities version, or the Newtonian limit, we have seen above that gravity
manifests itself as the attractive force per unit mass, {\bf g}, of eqn. (9) that
pulls any massive test body present at a given point in space towards the body $W$
that originates the field. Since we assumed the Einstein LLI principle and from this
derived the WEP, this, together with the very natural UA of invariance in the laws of
physics throughout universal spacetime, lead us to the Einstein SEP which necessarily
implies the spacetime bending representation of the generalized gravitational
field [6,8].

A simple thought experiment (Einstein's lift) immediately shows that light rays
propagate along geodesics, and more specifically along null geodesics [8,14,26]. The
spacetime bending is dramatically evident when a light ray goes from one side to the
other of the freely falling elevator. For the observer attached to the elevator's
frame that indeed acts as a LLI frame, the light ray propagates in a straight line
from one side to the other of the elevator. But for the stationary observer who sees
the elevator falling with acceleration {\bf g}, the light ray bends along a path that
locally is seen as a parabolic curve. Undoubtedly the most natural explanation for
the stationary observer is that spacetime bends and therefore the association that
this bending is a manifestation of the gravitational field of $W$, or rather that
this bending of spacetime is the gravitational field itself [26].   

Starting from the above fact taken as a given, the various metric theories proceed
from there to formulate their equations. In the version of Brans and Dicke a
scalar-tensor field is assumed. In Einstein's GR only a tensor
field is proposed. Following this maximally simplistic proposal and guided by general
considerations of general covariance (the need of arbitrary coordinates and tensor
laws) Einstein was led to his field equations in the presence of matter. And then GR
naturally unfolded [6,8,14,26].

We have shown here that our inertia proposal of [1] leads us, when limited by
the LLI principle, to the metric theories and therefore that it is consistent with
those theories and in particular with Einstein's GR. In addition, there is the
following interesting feature of our proposal.

From our analysis in [1], and in particular in Appendix B of [1], it was
made clear that within the quantum vacuum inertia hypothesis
there proposed, the mass of the object, $m$, could be viewed as the energy in the
equivalent vacuum electromagnetic zero-point field captured within the structure of
the object and that readily interacted with the object. This
view properly and accurately matched with the complementary view, thoroughly
exposed in other parts of the paper [1], that presented inertia as the result of a
vacuum reaction effect, a kind of drag force exerted by the vacuum field on
accelerated objects. Quantitatively, both approaches lead to exactly the same inertial
mass and moreover they were partly complementary. Both viewpoints were needed. One
could not exist without the other. They were the two sides of the same coin.

The question is now why massive objects, when freely falling, also follow geodesic
paths. The tempting view suggested here is that, as massive bodies (according to
our analysis of [1], Appendix B) have a mass that is made of the vacuum
electromagnetic energy contained within their structure and that readily interacts
with such structure, it is no surprise that geodesics are their natural path of
motion during free fall. Electromagnetic radiation has been shown by Einstein to
follow precisely geodesic paths. The only difference now is that, as the radiation
stays within the accelerated body structure and is contained within that structure
and thereby its energy center moves subrelativistically, these geodesics are just
time-like and not null ones as in the case of freely propagating light rays.

We illustrate this with an example. Imagine a freely-falling electromagnetic cavity
with perfectly reflecting walls of negligible weight, so that all the weight is due
to the enclosed radiation. A simple plane wave mode decomposition shows that although
individual wavetrains do still move at the speed of light, the center of energy of
the radiation inside the cavity moves subrelativistically as the wavetrains reflect
back and forth. The wavetrains do indeed propagate along null geodesics, but the
center of energy propagates only along a time-like geodesic.

Neither our approach nor the conventional presentations of GR for
that matter, can offer a physical explanation of the mechanism of the bending of
spacetime as related to energy density. Misner, Thorne and Wheeler [14] present six
different proposed explanations. The sixth is the one we already mentioned due
to Sakharov [12,13] which starts from general vacuum considerations. As our approach
starts also from vacuum considerations, it naturally fits better the concept of the
conjecture of Sakharov [12] and Zeldovich [13] than the other proposals but it is not
inconsistent with any of them. In particular the strictly formal proposal of Hilbert
[14] that introduces the so-called Einstein-Hilbert Action, is also at the origin of
the Sakharov proposal. We plan to devote more work to exploring the connection of our
inertia [1,2] and gravity approach to the approach proposed by Sakharov [12]. This we
leave for a future publication.

\bigskip\noindent
9. CONCLUSIONS
\bigskip\noindent

The principal conclusions of this paper are:

(1) {\it Identity of inertial mass with gravitational mass, $m_i=m_g$.} It has been
shown that the approach of [1] contains this peculiar feature that, so far and
as we know, has never been explained. Rindler [7] calls this feature ``a very
mysterious fact,'' as indeed it has been up to the present. We expect with this work
to have shed some light on this peculiar feature.

(2) {\it Consistency of the quantum vacuum inertia hypothesis with Einstein's GR.} We
have already commented above, in particular in $\S 7$, on this interesting feature
presented in $\S 3$ that puts the vacuum inertia approach of [1] within the
mainstream thought of contemporary gravitational theories, specifically within that
of theories of the metric type and in particular in agreement with GR. The
full GR development of gravitation within the scope of the quantum vacuum inertia
hypothesis is presented in [7].

(3) {\it Newton's gravitational law from the quantum vacuum inertia hypothesis.}
By means of a simple argument based on potential theory we show how to obtain in a
natural way Newton's inverse square force with distance from our vacuum approach to
inertia of [1]. The simplicity of our approach contrasts with previous atempts to
accomplish this within the framework of SED theory [15,16,17,18,19].

(4) {\it Origin of weight and a physical mechanism to enforce motion along geodesic
trajectories for freely-falling objects}.
We have 
shown how this approach to inertia answers a fundamental question left open
within GR, viz. is there a physical mechanism that generates the inertia reaction
when non-geodesic motion is imposed on an object and which can manifest specifically
as weight. Or put another way, while geometrodynamics dictates the spacetime metric
and thus specifies geodesics, is there an identifiable mechanism for enforcing motion
along geodesic trajectories? The quantum vacuum inertia hypothesis represents a
significant first step in providing such a mechanism.

\bigskip\noindent
ACKNOWLEDGEMENTS
\bigskip\noindent
We thank D. C. Cole, Y. Dobyns and M. Ibison for interesting and useful discussions.
AR received partial support from the California Institute for Physics and
Astrophysics via a grant to Cal. State Univ. at Long Beach. This work is based upon a
study carried out under NASA contract NASW-5050.

\bigskip\noindent
REFERENCES
\bigskip\noindent

\newcount\q \q=0
\def\nref {\global\advance\q by1 \item{\the\q.}}

\nref A. Rueda and B. Haisch, Found. Physics. 28, 1057 (1998).

\nref A. Rueda and B. Haisch, Phys. Lett. A 240 (1998) 115. This is a summary of the
main results of [1].

\nref B. Haisch, A. Rueda  and H.E. Puthoff, Phys. Rev A 48
(1994) 678. This was the first work in a search for the origin of inertia in the
quantum vacuum. The fact that it dealt with a too concrete model and that the
development was mathematically involved led to the necessity of showing that if this
proposal for inertia is true there must  exist an asymmetry in the electromagnetic
vacuum when seen from the standpoint of an accelerated observer. This last was the
work reported in Refs. [1] and [2] above.

\nref L. de La Pe\~na and A.M. Cetto, The
Quantum Dice -Ð An introduction to Stochastic Electrodynamics. (Kluwer Acad. Publ.,
Fundamental Theories of Physics Series, Dordrecht, Holland, 1996) and references
therein.

\nref T. H. Boyer, Phys. Rev. D, 11, 790 (1975).

\nref C.W. Will, Theory and Experiment in Gravitational Physics (Cambridge University
Press, Cambridge, 1993) pp 22--24. 

\nref R. Tung, B. Haisch and A. Rueda, companion paper.

\nref  W. Rindler, Essential Relativity Ð Special, General
and Cosmological (Springer Verlag, Heidelberg, 1977), p. 17.

\nref T. H. Boyer, Phys. Rev. D, 21, 2137 (1980) and Phys. Rev. D, 29, 1089 (1984).

\nref A. Einstein, Ann. Phys. 35, 898 (1911). For a translation see C.W. Kilmister,
General Theory of Relativity (Pergamon, Oxford, 1973), pp. 129--139. 

\nref See,e.g., P.A.M. Dirac, Directions in Physics (Wiley, New York, 1978), in
particular Section 5,  ``Cosmology and the gravitational constant,'' pg. 71 ff.
 
\nref A.D. Sakharov, Doklady Akad. Nauk S.S.S.R. 177 70-71 (1967) (English translation
in Sov. Phys. Doklady 12, 1040-1041  (1968)) 

\nref Yu. B. Zeldovich. Zh. Eksp. \& Teor.
Fiz. Pis'ma 6, 883-884 (1967) (English translation in Sov. Phys. -- JETP Lett. 6,
316-317 (1967))

\nref C.W. Misner, K.S. Thorne and J.A. Wheeler, Gravitation (Freeman,
New York, 1971) pp 426--428.

\nref H.E. Puthoff, Phys. Rev. A 39, 2333 (1989).

\nref S. Carlip, Phys. Rev A 47, 3452 (1993).

\nref H.E. Puthoff, Phys. Rev. A 47, 3454 (1993).

\nref B. Haisch, A. Rueda and H.E. Puthoff, Spec. Science and Technology 20, 99
(1997).

\nref D. C. Cole, A. Rueda and K. Danley, Phys. Rev. A, 63, OS4101
(2001).

\nref H.E. Puthoff, ``Polarizable-Vacuum (PV) representation of general relativity,''
Institute for Advanced Studies at Austin, preprint (1999).

\nref R. H. Dicke,
``Gravitation without a principle of equivalence,'' Rev. Mod. Phys. 29, 363-376
(1957). See also R.H. Dicke, ``Mach's Principle and Equivalence,'' in Proceedings of
the International School of Physics ``Enrico Fermi'' Course XX, Evidence for
Gravitational Theories, ed. C. Moller (Acad. Press, New York, 1961), pp 1--49.

\nref H.A. Wilson, Phys. Rev. 17, 54-59 (1921).

\nref Our arguments will partially be based on potential theory, see, e.g. O.D.
Kellog, Foundations of Potential Theory (Dover, New York, 1953) pp 34-39, in
particular see Ex 3, p. 37.

\nref A more detailed exposition of this and several other points related to this
argument in a more scholarly and pedagogical vein is in preparation: D. C. Cole, A.
Rueda, B. Haisch (2001).

\nref Y. Dobyns, A. Rueda and B. Haisch, Found. Phys., 30, (1), 59, (2000).

\nref See. e.g., R.M. Wald, General Relativity (Univ. of Chicago Press, Chicago, 1984)
pg. 67; and for a more popularizing account, R.M. Wald, Space, Time and Gravity Second
Edition (Univ. of Chicago Press, Chicago, 1992) Ch. 3 and in particular pp 33--34.

\nref B. Haisch and A. Rueda, Phys. Lett. A, 268, 224 (2000).

\vfill\eject
\centerline{TABLE 1.}
\centerline{Comparison of Standard View of Mass and Quantum Vacuum Inertia
Hypothesis}
\bigskip
\hrule\smallskip\hrule\medskip

\vbox{\settabs2\columns
\+ Standard  View of Mass& Quantum Vacuum Inertia Hypothesis
\cr}
\medskip\hrule
\bigskip\centerline{INERTIA REACTION FORCE}\bigskip

\vbox{\settabs2\columns
\+ accelerating object experiences event horizon
&  accelerating object experiences event horizon\cr\medskip

\+ event horizon promotes some quantum vacuum
&  event horizon promotes some quantum vacuum\cr
\+ \qquad (QV) energy to ``real photons''
&  \qquad (QV) energy to ``real photons''\cr\medskip

\+ accelerating object experiences QV ``real photons''
&  accelerating object experiences QV ``real photons''\cr
\+ \qquad as Unruh-Davies radiation
&  \qquad as Unruh-Davies radiation {\it plus Rindler flux}\cr\medskip

\+ Higgs field can generate mass-energy for quarks
&  Higgs field can generate mass-energy for quarks\cr
\+ \qquad and electrons
&  \qquad and electrons\cr\medskip

\+ inertia reaction force arises from intrinsic 
&  inertia reaction force is an acceleration-dependent\cr
\+ \qquad property of matter
&  \qquad drag force resulting from the Rindler flux
\cr\medskip

\+ {\bf f}=m{\bf a} is postulated
& {\bf f}=m{\bf a} ensues from hypothesis\cr\medskip}

\bigskip\centerline{GRAVITATIONAL FORCE/WEIGHT}\bigskip

\vbox{\settabs2\columns

\+ gravitating body determines geodesics
&  gravitating body determines geodesics\cr\medskip

\+ light rays and freely-falling objects follow geodesics
& light rays and freely-falling objects follow geodesics \cr\medskip

\+ weight of stationary object is an inertia reaction
&  weight of stationary object is an inertia reaction\cr
\+ \qquad force due to deviation from geodesic
& \qquad force due to deviation from geodesic\cr\medskip

\+ inertia reaction force arises from intrinsic
&  inertia reaction force is a metric-dependent drag \cr
\+ \qquad property of matter
&  \qquad force resulting from the Rindler flux
\cr\medskip

\+ $m_{inertial} = m_{gravitational}$ is postulated
&  $m_{inertial} = m_{gravitational}$ ensues from hypothesis \cr
}

\medskip\hrule
\bye